\documentclass[10pt,conference,letterpaper]{IEEEtran}
\usepackage{amsthm}
\usepackage{epsfig}
\usepackage{xfrac}
\usepackage{amsmath}
\usepackage{cases}
\usepackage{amssymb,mathrsfs,dsfont}
\usepackage{enumerate}
\usepackage{color}
\usepackage{textcomp}
\usepackage{tikz}
\usepackage{pgfplots}
\usepackage{mathtools}
\usepackage{scrextend}
\newtheorem{theorem}{Theorem}
\newtheorem{lemma}{Lemma}

\newtheorem{corollary}{Corollary}

\newtheorem{note}{Note}

\usetikzlibrary{dsp,chains}
\usetikzlibrary{shapes,arrows}
\usetikzlibrary{positioning,shapes,shadows,arrows,scopes,decorations}
\usetikzlibrary{shapes.multipart}

\def\md{\mathbb}

\def\eps{\varepsilon}
\def\tn{\textnormal}
\def\wt{\widetilde}

\def\Expt{\md{E}}

\def\db{\mathrm{dB}}
\def\latticeN{N}

\def\NL{{N_\Lambda}}
\def\nL{{n_\Lambda}}
\def\NC{{N}}
\def\nC{{n}}
\def\pairind{p}
\def\V0{\mathcal{V}_0}

\def\dotleq{\stackrel{.}{\leq}}
\def\dotgeq{\stackrel{.}{\geq}}

\def\FBa{{\bM{a}}}

\newcommand{\bM}[1]{\boldsymbol{#1}}

\newcommand{\dfn}{ \stackrel{\tn{def}}{=} }

\def\p2p{point-to-point}

\newcommand{\moduloLattice}[1] {\mathbb{M}_{\Lambda}\left[#1\right]}
\newcommand{\quantLattice}[1] {\mathbb{Q}_{\Lambda}\left[#1\right]}

\newcommand{\Pe}{p_e}
\newcommand{\Pmod}{p_\mathrm{mod}}

\newcommand{\latticeLoose}{L}
\newcommand{\BCcov}{\Sigma}
\newcommand{\BCcoveq}{\Sigma_{\mathrm{eq}}}
\newcommand{\Peq}{P_{\mathrm{eq}}}

\newcommand{\zcorr}{r}
\newcommand{\ShannonC}[1]{{\tfrac{1}{2}\log\left(1+ #1\right)}}
\newcommand{\codebookpart}{C}
\newcommand{\feedbackpart}{L}
\IEEEoverridecommandlockouts

\usepackage[noadjust]{cite}

\begin{document}
\title{The AWGN BC with MAC Feedback: A Reduction to Noiseless Feedback via Interaction}
\author{Assaf~Ben-Yishai and Ofer~Shayevitz
\thanks{The authors are with the Department of EE--Systems, Tel Aviv University, Tel Aviv, Israel \{assafbster@gmail.com, ofersha@eng.tau.ac.il\}. 
This work was supported by the Israel Science Foundation under grant agreement no. 1367/14, and by the Marie Curie Career Integration Grant (CIG) under grant agreement no. 631983.}
}


\maketitle

\begin{abstract}
We consider the problem of communication over a two-user Additive White Gaussian Noise Broadcast Channel (AWGN-BC) with an AWGN Multiple Access (MAC) active feedback. We describe a constructive reduction from this setup to the  well-studied setup of linear-feedback coding over the AWGN-BC with noiseless feedback (and different parameters). This reduction facilitates the design of linear-feedback coding schemes in the (passive) noiseless feedback regime, which can then be easily and constructively transformed into coding schemes in the MAC feedback regime that attain the exact same rates. Our construction introduces an element of interaction into the coding protocol, and is based on modulo-lattice operations. As an example, we apply our method to the Ozarow-Leung scheme, and demonstrate how MAC feedback can be used to enlarge the capacity region of the AWGN-BC. 
\end{abstract}

\section{Introduction}
It is well known that feedback can enlarge the capacity region of a non-degraded AWGN-BC, yet the capacity region with feedback generally remains unknown  \cite{ElGamalKimBook}. In \cite{OzarowBC}, Ozarow and Leung (OL) introduced a coding scheme that attains rate pairs outside the non-feedback capacity region, and also showed that their scheme is not optimal. In \cite{LQGbroadcast}, Ardestanizadeh, Minero and Franceschetti used the LQG approach from stochastic control theory to derive feedback coding schemes that exceed the OL sum rate. More recently, Amor, Steinberg and Wigger \cite{FB_MAC_BC_duality} characterized the capacity region with uncorrelated noises and Linear Feedback Coding (LFC), and showed that the LQG maximizes the sum-rate for the symmetric uncorrelated AWGN-BC among all LFC schemes. 

The case of the AWGN-BC with noisy feedback was studied by several authors using various feedback noise models \cite{shayevitzWigger,micheleNoisyBC,venkataramanan2013achievable}. This paper also considers the case of noise in the feedback, but is conceptually different from previous works in two important aspects. First, we assume that the feedback of the AWGN-BC channel is AWGN-MAC, a more practical assumption for wireless networks models. More importantly, we describe a constructive reduction from this setup to the well-studied setup of LFC schemes over the AWGN-BC with noiseless feedback (and different parameters). This reduction facilitates the design of LFC schemes in the (passive) noiseless feedback regime, which can then be easily and constructively transformed into coding schemes in the AWGN-MAC feedback regime that attain the exact same rates. Our construction introduces an element of interaction into the coding protocol, and is based on modulo-lattice operations. This approach is an extension of our previous work on noisy feedback for the point-to-point AWGN channel \cite{SimpleInteractionAllerton2014}\cite{GaussianInteractionISIT}. As a proof of concept, we apply our method to the OL scheme and demonstrate how AWGN-MAC feedback can be used to increase the capacity region of the AWGN-BC.  

\section{Preliminaries}
We write $\log$ for base $2$ logarithm, and $\ln$ for the natural logarithm. Vectors are written in boldface (e.g. $\bM{x}$), and superscripts are used to emphasize the vector length if necessary, e.g., $\bM{x}^n\dfn [x_1,\ldots,x_n]$. We write $a_n\dotgeq b_n$ to mean  $\liminf_{n\rightarrow\infty}\frac{1}{n}\ln\left(\frac{a_n}{b_n}\right)\geq 0$, and similarly define $\dotleq$ and $\doteq$. 

\subsection{Lattice Definitions and Properties}
\label{sec:lattice-properties}
\begin{itemize}
\item A lattice of dimension $\latticeN$ is denoted by $\Lambda=G\cdot\mathbb{Z}^\latticeN$ where $G$ is the generating matrix. 
\item $V(\Lambda)=|\det(G)|$ is the lattice cell volume.
\item The nearest neighbor quantization of $\bM{x}$ w.r.t. the lattice $\Lambda$ is denoted by $\quantLattice{\bM{x}}$.
\item The fundamental (Voronoi) cell of $\Lambda$ is denoted by $\V0=\{\bM{x}:\quantLattice{\bM{x}}=\bM{0} \}$.
\item The Modulo-$\Lambda$ operation is $\moduloLattice{\bM{x}}\dfn \bM{x}-\quantLattice{\bM{x
}}$.
\item $\moduloLattice{\cdot}$ satisfies the \textit{distributive law} : $\moduloLattice{\moduloLattice{\bM{x}}+\bM{y}}=\moduloLattice{\bM{x}+\bM{y}}$.
\item The volume to noise ratio (VNR) of a lattice in the presence of AWGN with variance $\sigma^2$ is $\mu\dfn V^{2/\latticeN}(\Lambda)/\sigma^2$. 
\item The normalized second moment of a lattice $\Lambda$ is  $G(\Lambda)\dfn \sigma^2(\Lambda)/V^{2/\latticeN}(\Lambda)$, where $\sigma^2(\Lambda)=\frac{1}{\latticeN}\Expt{(\|\bM{U} \|^2)}$ and $\bM{U}$ is uniformly distributed on $\V0$.
\end{itemize}

\begin{lemma}\label{lemma:joint-source-channel-1}
For an i.i.d Gaussian vector $\bM{x}$ of size $\NL$ whose elements have zero mean and variance $\sigma^2$, there exists lattices $\Lambda$ with size $\NL$ and second moment $\sigma^2(\Lambda)=L\cdot\sigma^2$ for which 
\begin{align}
\label{eq:pmod}
\Pmod\dfn \Pr(\bM{x}\notin \V0) \dotleq e^{-\NL E_p(\latticeLoose)}
\end{align}
where  $E_p(\cdot)$ is the Poltirev error exponent given by \cite{RamiLattices,GoodLattices}:
\begin{align*}
E_p(x) = 
\begin{cases}
\frac{1}{2}\left(x-1-\ln (x)\right) & \text{if } 1<x\leq 2 \\
\frac{1}{2}\left(\ln(x)+\ln(\frac{e}{4}) \right) & \text{if } 2<x\leq 4 \\
\frac{1}{8}x & \text{if } x>4 \\
\end{cases}
\end{align*}
\end{lemma}

For the sake of analysis in this work, it is enough to state that $E_p(x)>0$ for $x>1$.
The proof of Lemma~\ref{lemma:joint-source-channel-1} is based on the existence of lattices that are good for both channel coding and source coding as shown in  \cite[Theorem 5]{GoodLattices}. A similar statement was previously given in \cite{KochmanZamirJointWZWDP}.

\section{Setup}
\label{sec:setup}
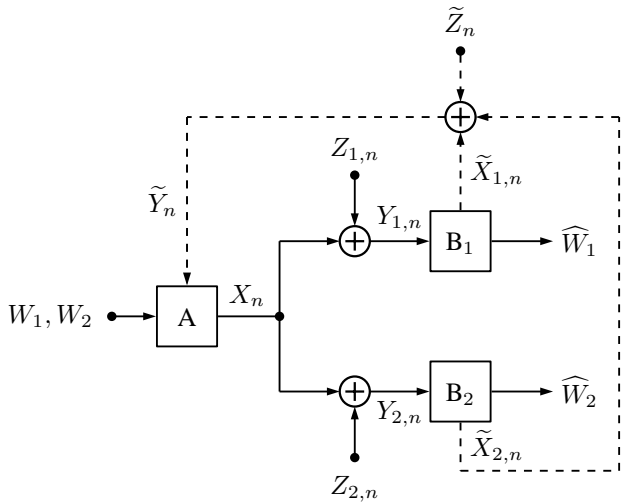
\begin{figure}
\centering
\newif\ifTIKZstandalone
\TIKZstandalonefalse

\ifTIKZstandalone
\documentclass[10pt,journal,twocolumn]{IEEEtran}
\usepackage{amsmath}
\usepackage{amssymb,mathrsfs,dsfont}
\usepackage{tikz}
\usetikzlibrary{dsp,chains}
\usetikzlibrary{shapes,arrows}
\usetikzlibrary{positioning,shapes,shadows,arrows,scopes,decorations}
\usetikzlibrary{shapes.multipart}

\begin{document}
\fi

\begin{tikzpicture}

\node[dspnodefull,dsp/label=left] (terAin){$W_1,W_2$};
\node[dspsquare,right of = terAin] (terA) {A}; 
\draw[dspconn] (terAin) -- (terA);
\node[dspnodefull,right of = terA,node distance=3.5em] (terAout) {};
\draw[dspline] (terA)  edge node [above] {$X_n$} (terAout) ;

\node[coordinate,above of = terAout] (Adder1Left) {};
\draw[dspline] (terAout) -- (Adder1Left);
\node[dspadder,right of = Adder1Left] (Adder1) {}; 
\draw[dspconn] (Adder1Left)--(Adder1);
\node[dspnodefull,above of = Adder1,dsp/label=above,node distance = 2.5em] (Z1) {$Z_{1,n}$}; 
\draw[dspconn] (Z1)--(Adder1);
\node[dspsquare,right of = Adder1,node distance=4em] (terB1) {$\text{B}_1$}; 
\draw[dspconn] (Adder1) edge node [above] {$Y_{1,n}$} (terB1);
\node[coordinate,right of = terB1,node distance = 3.5em] (W1){};
\draw[dspconn] (terB1)--(W1) ;
\node[right of = W1,node distance = 1em] () {$\widehat{W}_1$};

\node[coordinate,below of = terAout] (Adder2Left) {};
\draw[dspline] (terAout) -- (Adder2Left);
\node[dspadder,right of = Adder2Left] (Adder2) {}; 
\draw[dspconn] (Adder2Left)--(Adder2);
\node[dspnodefull,below of = Adder2,dsp/label=below,node distance = 2.5em] (Z2) {$Z_{2,n}$}; 
\draw[dspconn] (Z2)--(Adder2);
\node[dspsquare,right of = Adder2,node distance=4em] (terB2) {$\text{B}_2$}; 
\draw[dspconn] (Adder2) edge node [below] {$Y_{2,n}$} (terB2);
\node[coordinate,right of = terB2,node distance = 3.5em] (W2){};
\draw[dspconn] (terB2)  -- (W2);
\node[right of = W2,node distance = 1em] () {$\widehat{W}_2$};

\node[dspadder,above of = terB1, node distance = 4.7em] (MACsum)  {};
\draw[dspconn,dashed] (terB1) edge node [right] {$\widetilde{X}_{1,n}$} (MACsum);
\coordinate[] (aboveA) at (terA |- MACsum) {};

\draw[dspline,dashed] (MACsum) -- (aboveA);
\draw[dashed] (aboveA) edge node [left] {$\widetilde{Y}_{n}$} (terA);
\draw[dspconn,dashed] (aboveA) -- (terA);

\node[coordinate, below of = terB2,node distance = 3em] (terB2above) {};
\draw[dspline,dashed] (terB2) edge node [right] {$\widetilde{X}_{2,n}$} (terB2above);
\node[coordinate, right of = terB2above,node distance = 6em] (B2_2) {};
\draw[dspline,dashed] (terB2above)--(B2_2);
\node[coordinate] (rightMACsum) at (B2_2 |- MACsum) {};
\draw[dspline,dashed] (B2_2)--(rightMACsum);
\draw[dspconn,dashed] (rightMACsum)--(MACsum);

\node[dspnodefull,above of = MACsum,dsp/label=above,node distance = 2.5em] (Zt) {$\widetilde{Z}_n$};
\draw[dspconn,dashed] (Zt)--(MACsum);
\node[below of = Z2,node distance = 0.5em] () {}; 
\end{tikzpicture}

\ifTIKZstandalone
\end{document}
\fi
\caption{\label{fig:blockBC_MAC} A block diagram of BC with MAC feedback.}
\end{figure}
Our setup is depicted in Fig.~\ref{fig:blockBC_MAC} and is defined as follows. Terminal A is connected to both Terminal $\text{B}_1$ and $\text{B}_2$ through an AWGN-BC. The channel input from Terminal A at time $n$ is $X_n$ and the output to Terminal $\text{B}_i$ is $Y_{i,n}$ for $i\in\{1,2\}$. The input-to-output relation is given by:
 \begin{align*}
 Y_{i,n}=X_n+Z_{i,n} \text{ for }i\in\{1,2\}
 \end{align*}
The noise pairs $(Z_{1,n},Z_{2,n})$ are Gaussian and independent between time instances, with zero mean and a covariance matrix
\begin{align*}
\BCcov\dfn\
\left[
\begin{matrix}
\sigma^2_1&\zcorr\sigma_1\sigma_2\\ \zcorr\sigma_1\sigma_2&\sigma^2_2
\end{matrix}
\right]
\end{align*}

The feedback link is an AWGN-MAC, whose input from Terminal $\text{B}_i$ at time $n$ is $\wt{X}_{i,n}$ and the corresponding output at Terminal A is $\wt{Y}_n$. The input-to-output relation is given by 
 \begin{align*}
 \wt{Y}_{n}=\wt{X}_{1,n}+\wt{X}_{2,n}+\wt{Z}_n
 \end{align*}
The noise process $\{\wt{Z}_n\}$ is i.i.d zero mean Gaussian with variance is $\Expt{\wt{Z}_{n}^2}=\wt{\sigma}^2$, and is independent of the feedforward noise process.

Terminal A is in possession of a pair of independent messages  $W_1\sim \textrm{Uniform}([M_1])$ and $W_2\sim \textrm{Uniform}([M_2])$, to be described to Terminals $\text{B}_1$ and $\text{B}_2$ respectively over $N$ rounds of communication. To that end, the terminals can employ an interactive scheme defined by a three functions $(\varphi,\wt{\varphi}_1,\wt{\varphi}_2)$ as follows: At time $n$, Terminal A sends a function of its message pair $(W_1,W_2)$ and possibly of past feedback channel outputs over the feedforward channel, i.e., 
\begin{align}
\label{eq:feedbackcoding}
  X_n=\varphi_n(W_1,W_2,\wt{Y}^{n-1}).
\end{align}
Similarly, Terminals $\text{B}_1$ and $\text{B}_2$ send functions of their past observations to Terminal A over the feedback channel, i.e., 
\begin{align*}
  \wt{X}_{i,n}=\wt{\varphi}_{i,n}(Y_i^n)\text{ for }i\in\{1,2\}
\end{align*}
We also note that, in general, we allow these functions to further depend on common randomness shared by the terminals.

As for power constraints, we assume that Terminal A is subject to $\sum_{n=1}^N\mathbb{E}X_n^2\leq N\cdot P$ and Terminals $\text{B}_i$ are subject to identical power constraints $\sum_{n=1}^N\mathbb{E}\wt{X}_{i,n}^2 \leq N\cdot \wt{P}$ for $i\in\{1,2\}$. 

An interactive scheme $(\varphi,\wt{\varphi}_1,\wt{\varphi}_2)$ is associated with a rate pair $R_i\dfn \frac{\log{M_i}}{N}$ (for $i\in\{1,2\}$) and an error probability $\Pe(N,R_1,R_2)$, which is the probability that at least one of the Terminals $\text{B}_i$ errs in decoding its designated message $W_i$ at time $N$, under the optimal decision rule. 

Note that in the classical noiseless (and passive) feedback AWGN-BC setup \cite{OzarowBC}, Terminal A sees at time $n$ both $Y_{1,n-1}$ and $Y_{2,n-1}$ and can use both for coding. Namely: $ X_n=\varphi_n(W_1,W_2,{Y}_1^{n-1},{Y}_2^{n-1})$. We refer to this setting as \textit{noiseless feedback}.

\section{Coding Schemes}
\subsection{LFC Schemes}
We consider LFC schemes as defined in \cite{FB_MAC_BC_duality}. In such schemes the transmission functions \eqref{eq:feedbackcoding} reduce to:
\begin{align}
\label{eq:LFCY}
X_n = {\codebookpart}_n-\sum_{i=1}^2{\feedbackpart}_{i,n}
\end{align}
where ${\codebookpart}_n={\codebookpart}_n(W_1,W_2)$ is the \textit{codebook element}, i.e. a function of the messages, but not the feedback. The \textit{feedback elements}, ${\feedbackpart}_{i,n}={\feedbackpart}_{i,n}(\bM{Y}_i^{n-1})$, are linear combinations of the channel outputs: ${\feedbackpart}_{i,n}={\bM{a}}_{i,n}\bM{Y}_{i}^{n-1}$, and ${\bM{a}}_{i,n}$ are row vectors of length $n-1$. 

By a simple induction argument, it can be easily shown that \eqref{eq:LFCY} admits the following equivalent formulation:
\begin{align}
\label{eq:XcleanFB}
X_n = \overline{\codebookpart}_n-\sum_{i=1}^2\overline{\feedbackpart}_{i,n},
\end{align}
where $\overline{\codebookpart}_n=\overline{\codebookpart}_n(W_1,W_2)$ and $\overline{\feedbackpart}_{i,n}=\overline{\bM{a}}_{i,n}\bM{Z}_{i}^{n-1}$. Namely, the linear combinations of the outputs can be replaced with (different) linear combinations of the noises, by properly modifying the codebook elements. 

Observe that in \eqref{eq:XcleanFB}, the codebook element $\overline{\codebookpart}_n$ is statistically independent of the feedback element $\sum_i \overline{\feedbackpart}_{i,n}$ and both have bounded power. This highlights the power splitting between the codebook and the feedback. In contrast, in \eqref{eq:LFCY} the codebook element  ${\codebookpart}_n$ is highly correlated with the feedback element $\sum_i {\feedbackpart}_{i,n}$, and their individual powers can grow unbounded (e.g. in OL where the powers grow exponentially). This can be interpreted by viewing $\sum_i {\feedbackpart}_{i,n}$ as a linear predictor for the codebook element ${\codebookpart}_n$ which facilitates the power savings at the transmitter. The typical exponential power growth corresponds to the ``zoom-in'' effect which is often observed in feedback communication.

It is instructive to note that the feedback elements in \eqref{eq:LFCY} are computable at Terminals $B_1$ and $B_2$, but typically cannot be fed back to Terminal $A$ under any finite power constraint. In contrast, the feedback elements in \eqref{eq:XcleanFB} are not computable at Terminals $B_1$ and $B_2$, yet have bounded powers. 

\subsection{Modulo-Lattice LFC (MLLFC) Schemes}
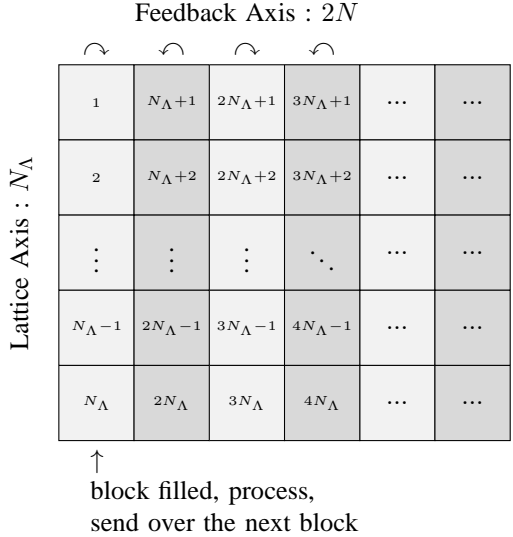
\begin{figure}
\centering
\newif\ifTIKZstandalone
\TIKZstandalonefalse

\ifTIKZstandalone
\documentclass[10pt,journal,twocolumn]{IEEEtran}
\usepackage{amsmath}
\usepackage{amssymb,mathrsfs,dsfont}
\usepackage{tikz}
\usetikzlibrary{dsp,chains}
\usetikzlibrary{shapes,arrows}
\usetikzlibrary{positioning,shapes,shadows,arrows,scopes,decorations}
\usetikzlibrary{shapes.multipart}

\begin{document}
\fi

\begin{tikzpicture}

\newcommand{\eval}[1]{\pgfmathparse{#1}\pgfmathresult}
\def\NL{{N_\Lambda}}
\definecolor{Gray1}{gray}{0.95}
\definecolor{Gray2}{gray}{0.85}
\newcommand{\smallfont}[1]{{\scriptscriptstyle #1}}

\def\xylab{{"$\smallfont{1}$","$\smallfont{\NL+1}$","$\smallfont{2\NL+1}$","$\smallfont{3\NL+1}$","$...$","$...$",
"$\smallfont{2}$","$\smallfont{\NL+2}$","$\smallfont{2\NL+2}$","$\smallfont{3\NL+2}$","$...$","$...$",
"$\vdots$","$\vdots$","$\vdots$","$\ddots$","$...$","$...$",
"$\smallfont{\NL-1}$","$\smallfont{2\NL-1}$","$\smallfont{3\NL-1}$","$\smallfont{4\NL-1}$","$...$","$...$",
"$\smallfont{\NL}$","$\smallfont{2\NL}$","$\smallfont{3\NL}$","$\smallfont{4\NL}$","$...$","$...$"}};

\foreach \ii in {0,2,4}
{
\foreach \jj in {0,...,4}
{
  \draw (\ii,-\jj) +(-0.5,-0.5) rectangle ++(0.5,0.5) [fill=Gray1];
  \draw (\ii,-\jj) node {\eval{\xylab[\ii+6*\jj]}};
}
}

\foreach \ii in {1,3,5}
{
\foreach \jj in {0,...,4}
{
  \draw (\ii,-\jj) +(-0.5,-0.5) rectangle ++(0.5,0.5) [fill=Gray2];
  \draw (\ii,-\jj) node {\eval{\xylab[\ii+6*\jj]}};
}
}

\draw (0,0.7) node {$\curvearrowright$};
\draw (2,0.7) node {$\curvearrowright$};
\draw (1,0.7) node {$\curvearrowleft$};
\draw (3,0.7) node {$\curvearrowleft$};
\draw (2,1.2) node {Feedback Axis : $2N$};
\draw (-1,-2.0) node [rotate=90] {Lattice Axis : $\NL$ };
\draw (1.7,-5.2) node  {
  \begin{tabular}{l}
    $\uparrow$ \\
    block filled, process, \\
    send over the next block
  \end{tabular}};

\end{tikzpicture}

\ifTIKZstandalone
\end{document}
\fi
\caption{\label{fig:blockcoding} Two dimensional coding for MLLFC}
\end{figure}

We are now ready to present the construction of the \textit{modulo lattice linear feedback coding} (MLLFC) schemes for the AWGN-BC with AWGN-MAC feedback. The method is an extension of the one described in \cite{GaussianInteractionISIT} for point-to-point AWGN with noisy feedback. We use $2\NL$ LFC schemes each of length $\NC$ interleaved in time, which can be thought of as a two-dimensional $\NL \times 2\NC$ coding. The notion of two dimensional coding is depicted in Fig.~\ref{fig:blockcoding}. We do lattice coding for the feedback over the vertical axis (where adjacent symbols have time difference of $1$), and use that feedback for LFC coding over the horizontal axis (where adjacent symbols have time difference of $2\NL$ ). The technical reason for the factor $2$ in $2\NL$ is to accommodate the inherent delay of $\NL$ of the lattice coding operation, and facilitate the use of feedback. 

Let us now describe the coding scheme in more detail. 
\begin{itemize}
\item We use an LFC with functions and parameters $\codebookpart_n$ and $\bM{a}_{i,n}$ according to formulation \eqref{eq:LFCY}, and  $\overline{\codebookpart}_n$ and $\overline{\bM{a}}_{i,n}$ according to formulation \eqref{eq:XcleanFB}.
\item We run multiple instances of the same LFC scheme described above. We index each instance by the index pair $(\nL,\pairind)$, where $\nL\in [\NL]$ and $\pairind\in\{0,1\}$ (the parity index). The feedforward of scheme $(\nL,\pairind)$ is sent over time instance 
$\nL+2\NL(\nC+\pairind)$ where $\nC\in[\NC]$.
\item The blocks of length $\NL$ are indexed by an index pair $(\nC,\pairind)$. The block corresponding to these indices pertains to time indices $1+\NL(\nC+\pairind-1)$ to $\NL(\nC+\pairind)$.
\item At the end of block $(\nC-1,\pairind)$, Terminals $\text{B}_1$ and $\text{B}_2$ each have 
the outputs corresponding to schemes $(1,\pairind)$ through $(\NL,\pairind)$ at step $\nC-1$.
For simplicity of exposition we denote these vectors by $\bM{Y}_{i}^{\nC-1}$ and omit the $\nL$ and $\pairind$ indices. Terminals $\text{B}_1$ and $\text{B}_2$ both compute the linear combinations $\FBa_{i,\nC}\bM{Y}_{i}^{\nC-1}$ for all $\NL$ schemes in this block. These $\NL$ elements are stacked in a vector denoted by ${\bM{\feedbackpart}}_{\i,\nC}$ (the index $\pairind$ is omitted for simplicity).
\item Over the following block, indexed $(\nC-1,\pairind+1)$, Terminals $\text{B}_i$ both send
\begin{align*}
\wt{\bM{X}}_{i,\nC-1}=\moduloLattice{\gamma_n {\bM{\feedbackpart}}_{i,\nC}+\bM{V}_{i,\nC}}
\end{align*}
where $\Lambda$ is a lattice of dimension $\NL$. $\moduloLattice{\cdot}$ is the lattice modulo operations. $\bM{V}_{i,\nC}$
 are dither variables, that are i.i.d and uniformly distributed over the lattice fundamental Voronoi cell $\V0$.
\item Terminal A receives the following vector:
\begin{align*}
\wt{\bM{Y}}_{\nC-1}=\sum_{i=1}^2\wt{\bM{X}}_{i,\nC-1}+\wt{\bM{Z}}_{\nC-1}
\end{align*}
It calculates the codebook elements $\codebookpart_n$ and $\overline{\codebookpart}_n$ for all $\NL$ schemes in the block and stacks them in a vector $\bM{\codebookpart}_{n}$ and  $\overline{\bM{\codebookpart}}_{n}$ respectively. 

\item At the following block, Terminal A calculates:
\begin{align}
\label{eq:Kn}
\bM{K}_{\nC}=\moduloLattice{\gamma_n\bM{\codebookpart}_{\nC}-\left(\wt{\bM{Y}}_{\nC-1}-\sum_{i=1}^2\bM{V}_{i,\nC}\right)-\gamma_n\overline{\bM{\codebookpart}}_{\nC}}
\end{align}
Due to the distributive law of the modulo operation 
\begin{align*}
\bM{K}_{\nC}=\moduloLattice{
\gamma_n\sum_{i=1}^2\overline{\bM{\feedbackpart}}_{i,\nC}-\wt{\bM{Z}}_{\nC-1}
}
\end{align*}
where $\overline{\bM{\feedbackpart}}_{i,\nC}$ is a vector containing the feedback elements of all $\NL$ schemes. 
Finally, we send 
\begin{align}
\label{eq:Xn}
\bM{X}_{\nC}=\gamma_n^{-1}\bM{K}_n+\overline{\bM{\codebookpart}}_{n}.
\end{align}

We refer to the event in which $\gamma_\nC\sum_{i}{\overline{\bM{\feedbackpart}}}_{i,\nC}-\wt{\bM{Z}}_{n-1}\notin\V0$, as a \textit{modulo aliasing error}. If this event does not occur then 
\begin{align*}
\bM{X}_{\nC}=\bM{\codebookpart}_{n}-\sum_{i=1}^2{\bM{\feedbackpart}_{i,n}}-\gamma_n^{-1}\wt{\bM{Z}}_{n-1}.
\end{align*}

Note that $\gamma_n\overline{\bM{\codebookpart}}_{\nC}$ is subtracted in \eqref{eq:Kn} (inside the modulo) and then added back in \eqref{eq:Xn}. This is done to assure that the modulo operand in \eqref{eq:Kn} is Gaussian in the coupled system (see below and also \cite{SimpleInteractionAllerton2014}), which is crucial for the error analysis in the sequel. It is also interesting to note that in the OL scheme $\overline{\bM{\codebookpart}}_{\nC}=0$ for all $n>2$.

\item Inspecting a single scheme in the block (and omitting the scheme indexing) we obtain
\begin{align}
\label{eq:Xreduction}
{X}_{\nC}={\codebookpart}_{n}-\sum_{i=1}^2{{\feedbackpart}_{i,n}}-\gamma_n^{-1}\wt{{Z}}_{n-1}
\end{align}
which is equivalent to the LFC transmission \eqref{eq:XcleanFB} with an additional noise element $\gamma_n^{-1}\wt{{Z}}_{n-1}$ 
\end{itemize}

It is appropriate to note that the feedback transmission described above is actually the \textit{analog Modulo Lattice Modulation} of Kochman and Zamir \cite{KochmanZamirJointWZWDP}, with the exception that the source-related computations are distributed between two terminals and added over MAC. The essence of the scheme is in \eqref{eq:Xreduction}, implying that if no modulo aliasing error occurs in the feedback, Terminal A can apply LFC transmission, without explicitly knowing the channel output elements required for its calculation. The difference between the clean feedback transmission in \eqref{eq:LFCY} and the noisy feedback transmission in \eqref{eq:Xreduction} is the existence of a feedback noise element $-\gamma_n^{-1}\wt{{Z}}_{n-1}$. This feedback noise element can be effectively added to the forward channel noise; we refer to this phenomenon as \textit{noise insertion}. In addition, the feedback noise element consumes part of the transmission power; we refer to this phenomenon as \textit{power loss}.

\section{Main Result}
By choosing an appropriate lattice $\Lambda$, properly setting scaling parameters $\gamma_n$ and the LFC functions defined by $\codebookpart_n,\feedbackpart_{i,n},\overline{\codebookpart}_n$ and $\overline{\feedbackpart}_{i,n}$ one can calculate a set of achievable rates for the MLLFC scheme. Nevertheless, a joint optimization of all the aforementioned parameters is very involved. In the following theorem, which is the main contribution of this work,  we introduce a simple setting of the parameters that reduces MLLFC to LFC with clean feedback, achieving the same rates.
\begin{theorem}
\label{thrm:BCreduction}
Let Setup 1 denote an AWGN-BC with AWGN-MAC feedback with paramters $P,\BCcov,\wt{P},\wt{\sigma}^2$ (and $\wt{P}>\wt{\sigma}^2$). Let Setup 2 denote an AWGN-BC with noiseless feedback, feedforward power constraint $\Peq$, and covariance matrix $\BCcoveq$, given by 
\begin{align}
\Peq&=P\left(1-\frac{\wt{\sigma}^2}{\wt{P}}\right) \label{eq:Peq}\\
\BCcoveq&=\BCcov+\wt{\sigma}^2\frac{P}{\wt{P}}
\left[
\begin{matrix}
1 & 1 \\
1 & 1
\end{matrix}
\right]
.
\end{align}
Then for any LFC with constant transmission power ($\Expt{X_n^2}=\Peq$) for Setup 2, there exists an MLLFC for Setup 1 achieving the same rates. 
\end{theorem}

\begin{proof}
The proof is based on the MLLFC construction described the previous section, with a judicious (but not necessarily optimal) setting of the parameters. Firstly, we choose a lattice $\Lambda$ having a normalized second moment $\sigma^2(\Lambda)=\wt{P}$, that is good in the sense of Lemma~\ref{lemma:joint-source-channel-1} (i.e. achieving $\Pmod$ in the lemma). We also set:
\begin{align}
\label{eq:gammandef}
\gamma_n=\sqrt{\tfrac{\wt{P}}{P}}.
\end{align} 

Let us now explicitly describe the choice of the functions $\codebookpart_n,\feedbackpart_{i,n},\overline{\codebookpart}_n,\overline{\feedbackpart}_{i,n}$ corresponding to the LFC for the AWGN-BC with noiseless feedback characterized by $\Peq$ and $\BCcoveq$. 
In order to claim that all the rates achievable in this noiseless feedback setup are also achievable in the original setup, we need to validate the following properties of the MLLFC:
\begin{enumerate}
\item The overall error probability goes to zero as $\NC\rightarrow\infty$ and $\NL\rightarrow\infty$.
\item The power constraints are not violated.
\item The equivalent channel model after the MLLFC transformation is faithful to Setup 2. 
\end{enumerate}

We start by evaluating the error probability. As shown in \cite[Lemma 1]{SimpleInteractionAllerton2014}, the overall error probability can be bounded by the sum of modulo aliasing errors and errors of the LFC schemes, where all the probabilities are evaluated in a \textit{Gaussian coupled system}, i.e. a system that applies no modulo operations and whose statistics are Gaussian. Denoting the overall error probability by $\Pe$ and the error probability of a linear feedback coding scheme of length $\NC$ by $\Pe^{\mathrm{lfb}}(\NC)$ and referring to \eqref{eq:pmod} yields:
\begin{align*}
\Pe\dotleq 2\NC e^{-\NL E_p(\latticeLoose)}+2\NL\Pe^{\mathrm{lfb}}(\NC)
\end{align*}
Where $\latticeLoose$ is defined in Lemma~\ref{lemma:joint-source-channel-1}.
In order to guarantee that $\Pe\rightarrow 
0$ we need to verify that $L>1$, which assures that $E_p(\latticeLoose)>0$. We also need to assume that (say)  $\Pe^{\mathrm{lfb}}(\NC)=o(\NC^{-1})$ for all achievable rate pairs on Setup 2 (which is not a restrictive assumption). Then we can set $\NC=\NL$ and take the limit $N\to\infty$.

We now proceed to verify that $\latticeLoose>1$. Observe that by construction definition of the LFC in \eqref{eq:XcleanFB}, the signals $\overline{\codebookpart}_{\nC}$ and $\overline{\feedbackpart}_{1,\nC}+\overline{\feedbackpart}_{2,\nC}$ are independent, hence
\begin{align*}
\Peq=\Expt{\overline{\codebookpart}_{\nC}^2}+\Expt{\left(\overline{\feedbackpart}_{1,\nC}+\overline{\feedbackpart}_{2,\nC}\right)^2}.
\end{align*}
Let us write $\Expt{\overline{\codebookpart}_n^2}=(1-\eta_\nC)\Peq$ and $\Expt{(\overline{\feedbackpart}_{1,\nC}+\overline{\feedbackpart}_{2,\nC})^2}=\eta_\nC\Peq$ where $\eta_\nC\in[0,1]$.
A modulo aliasing error will occur in \eqref{eq:Kn} if 
$\bM{v}_n\notin \V0$ where
 $\bM{v}_n\dfn \gamma_\nC\sum_{i}{\overline{\bM{\feedbackpart}}}_{i,\nC}-\wt{\bM{Z}}_{n-1}$. In the coupled system, all the elements $\bM{v}_n$ are Gaussian and i.i.d with variance
\begin{align*}
\gamma_n^2\eta_n\Peq+\wt{\sigma}^2=\eta_n\wt{P}+(1-\eta_n)\wt{\sigma}^2,
\end{align*}
where the equality is due to \eqref{eq:Peq} and \eqref{eq:gammandef}. Since we assumed $\wt{P}>\wt{\sigma}^2$, then $\gamma_n^2\eta_n\Peq+\wt{\sigma}^2\leq\wt{P}$ yielding $L\geq 1$.\footnote{In order to guarantee that $L>1$ we in fact need to set $\gamma=\sqrt{\wt{P}/P}-\eps$ with a arbitrarily small positive $\eps$. }

From \eqref{eq:Xreduction} we see that if no modulo error occurs, then 
\begin{align*}
\Expt{X_{\nC}^2}&=\Expt{\overline{\codebookpart}_{\nC}^2}+\Expt{\left(\overline{\feedbackpart}_{1,\nC}+\overline{\feedbackpart}_{2,\nC}\right)^2}+\gamma_n^{-2}\Expt{(\wt{Z}_{n-1})^2}\\
&=P\left(1-\frac{\wt{\sigma}^2}{\wt{P}}\right)+\frac{P}{\wt{P}}\wt{\sigma}^2=P.
\end{align*}
Hence setting $\Peq$ as specified obeys the original power constraint. The feedback noise term $-\gamma_n^{-1}\wt{Z}_{n-1}$ can be effectively added to the feedforward channel noise, modifying the channel covariance matrix from $\BCcov$ to $\BCcoveq$ as specified in the theorem.

\end{proof}
\begin{corollary}
The rate region achieved by the OL scheme for Setup 2 is also achievable for Setup 1, using the associated modulo-lattice OL scheme. 
\end{corollary}
\begin{proof}
Since the OL scheme is a LFC with constant transmission power \cite{OzarowBC}, the claim follows by virtue of Theorem~\ref{thrm:BCreduction}.
\end{proof}

\begin{note}
If $\frac{\wt{P}}{\wt{\sigma}^2}\rightarrow\infty$, then $\Peq\rightarrow P$ and $\BCcoveq\rightarrow\BCcov$ and the rate region collapses to the one achieved by LFC with constant transmission power in a noiseless feedback setup with $P$ and $\BCcov$.
\end{note}

\section{Example - A Modulo Lattice OL Scheme}

\begin{figure}
\input{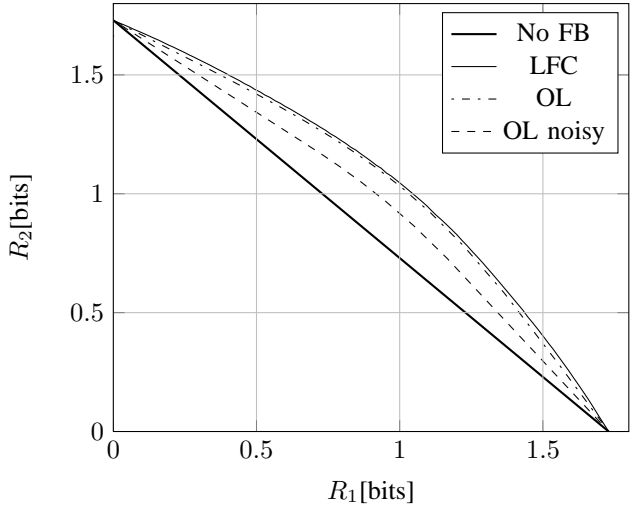}
\caption{\label{fig:BCFig1}
Achievable rates for BC without feedback, with noiseless feedback and optimal LFC, noiseless feedback and OL, and noisy feedback and OL after the transformation in Theorem~\ref{thrm:BCreduction}. The BC is symmetric and uncorrelated with parameters $P=10,\sigma^2_1=\sigma^2_2=1,r=0$, the MAC parameters are $\wt{P}=100,\wt{\sigma}^2=1$.}
\end{figure}


In this section we consider a symmetric AWGN-BC with independent noises and AWGN-MAC feedback, and juxtapose the rate region we achieve with the MLLFC obtained from the OL scheme, with some known results. For this setup, an inner bound is given by the capacity of AWGN-BC without feedback \cite{ElGamalKimBook}:
\begin{align*}
\bigcup_{\alpha\in[0,1]} \Big\{ R_1&\leq \ShannonC{\frac{\alpha P}{\sigma^2_1}},\\
R_2&\leq \ShannonC{\frac{(1-\alpha P)}{\sigma^2_2+\alpha P}}\Big\}.
\end{align*}

An outer bound on the LFC achievable region for the AWGN-BC LFC (with uncorrelated noises)
and noiseless feedback, is given by \cite{FB_MAC_BC_duality}
\begin{align*}
\mathcal{C}^{\mathrm{LFC}}_{\mathrm{BC}}\left(P,\BCcov\right) =
 \bigcup_{P'_1+P'_2=P} \mathcal{C}^{\mathrm{FB}}_{\mathrm{MAC}}\left(\frac{P'_1}{\sigma^2_1},\frac{P'_2}{\sigma^2_2}\right)
\end{align*}
where the right-hand-side corresponds to the capacity region of dual MAC problem with unit noise variance:
\begin{align*}
 \mathcal{C}^{\mathrm{FB}}_{\mathrm{MAC}}\left(P_1,P_2\right) \dfn& \\
\bigcup_{\rho\in[0,1]} \Big\{R_1\leq& \ShannonC{P_1(1-\rho^2)},  \\
R_2\leq& \ShannonC{P_2(1-\rho^2)},\\
R_1+R_2\leq& \ShannonC{P_1+P_2+2\sqrt{P_1P_2}\rho} \Big\}.
\end{align*}

A comparison between these bounds and our modulo-lattice OL scheme is given in Fig.~\ref{fig:BCFig1}. For reference, we also plotted the rate region of an OL scheme with noiseless feedback. The formulas for the rates of the OL scheme are more involved and appear in \cite{OzarowBC}. In the example of Fig.~\ref{fig:BCFig1}, all the noise variances are set to unity, i.e. $\sigma^2_1=\sigma^2_2=\wt{\sigma}^2=1$, and the noises are uncorrelated. The feedforward power is set to $P=10$ and the feedback power is set to be $10\db$ stronger, i.e. $\wt{P}=100$. The plotted OL region for noisy feedback is calculated by taking the convex hull of the union of the OL region after the transformation and the no feedback region. It is clear from the figure that the capacity region for the AWGN-BC with noisy AWGN-MAC feedback is strictly larger than the no feedback region.


\bibliographystyle{IEEEbib}
\bibliography{bibtex_references}

\end{document}